\documentclass[acmsmall,screen]{acmart}

\acmJournal{TWEB}
\acmVolume{0}
\acmNumber{0}
\acmArticle{0}
\acmYear{2026}
\acmMonth{8}
\acmDOI{XXXXXXX.XXXXXXX}

\usepackage{graphicx} 
\usepackage{url}
\usepackage{booktabs}
\usepackage{multirow}
\usepackage{subcaption}

\usepackage{todonotes}
\newcommand{\SC}[1]{}
\newcommand{\SB}[1]{}
\newcommand{\LC}[1]{}
\usepackage{tcolorbox}

\usepackage{courier}
\usepackage{paralist}

\newcommand{\revision}[1]{\textcolor{black}{#1}}

\begin{document}

\title{LLM-Based Website Classification for Targeted Web Measurements}

\author{Simone Bozzolan}
\orcid{0009-0009-1054-1234}
\affiliation{%
\institution{Università Ca' Foscari Venezia}
\city{Venice}
\country{Italy}}
\email{simone.bozzolan@unive.it}

\author{Stefano Calzavara}
\orcid{0000-0001-9179-8270}
\affiliation{%
\institution{Università Ca' Foscari Venezia}
\city{Venice}
\country{Italy}}
\email{stefano.calzavara@unive.it}

\author{Lorenzo Cazzaro}
\orcid{0000-0001-6479-2949}
\affiliation{%
\institution{University of Luxembourg}
\city{Esch-sur-Alzette}
\country{Luxembourg}}
\email{lorenzo.cazzaro@uni.lu}

\renewcommand{\shortauthors}{Bozzolan et al.}

\begin{abstract}
Web measurements based on top lists of popular websites are a well-established methodology for assessing the security and privacy landscape of the Internet. However, existing top lists are unlabeled and lack semantic information about the nature of the included websites, forcing researchers to rely on custom techniques to focus on specific website classes of interest. This makes \emph{targeted} web measurements, e.g., over shopping websites, challenging to carry out. In this paper, we investigate the use of Large Language Models (LLMs) to enable targeted web measurement studies. Building on prior literature, we identify key website classification tasks relevant to web measurements and highlight limitations in state-of-the-art classification approaches. We thus construct more carefully curated datasets to evaluate different LLMs on these tasks using standard performance measures. Our results show that LLMs can achieve strong performance across multiple classification scenarios, but the choice of the model and its configuration play a significant role in both the quality and the efficiency of the classification process. Finally, we use LLMs to emulate the setup of a targeted web measurement by automatically detecting adult websites in a subset of the Tranco list, without any need for custom techniques for dataset construction. Together, our benchmark evaluation and case study show that LLMs can enable high-quality targeted web measurements at scale directly over the Tranco list.
\end{abstract}

\ccsdesc[500]{Security and privacy~Software and application security}
\ccsdesc[300]{Information systems~World Wide Web}
\ccsdesc[300]{Computing methodologies~Artificial intelligence}

\keywords{web measurements, large language models}

\maketitle

\section{Introduction}


Web measurements are a popular tool to assess the current state of security and privacy on the Internet. Starting from a dataset of websites to analyze, web measurements leverage web crawling and automated analysis techniques to determine whether existing websites comply with security best practices~\cite{CalzavaraRB18,CalzavaraRR0S20}, suffer from known vulnerabilities~\cite{LekiesSJ13,SudhodananCCDAM17}, or are aligned with current privacy regulations~\cite{EnglehardtN16,MunirSIEST23}. Naturally, the representativeness and reliability of web measurements are only as good as the underlying datasets. Legacy work largely relied on public lists of popular websites (\emph{top lists}) created by private companies, e.g., the now discontinued Alexa ranking~\cite{daily-star_amazon_closing_alexa_2021}. However, these lists turned out to be brittle, unstable, and ultimately unreliable to draw meaningful conclusions, which motivated the creation of the Tranco ranking as a more robust alternative for security and privacy research~\cite{PochatGTKJ19}. Tranco aggregates multiple top lists to mitigate their bias and reduce popularity fluctuations, and it is now considered a reference dataset for web measurements, having been used in more than 600 academic publications.

Unfortunately, many web measurements cannot be meaningfully performed over Tranco as is, because Tranco is an \emph{unlabeled} dataset, with no semantic information beyond relative popularity. This makes \emph{targeted} web measurement studies challenging to carry out or fundamentally limited in practice. For example, prior work analyzed the privacy guarantees of the governmental website ecosystem~\cite{GotzeMISL22, SamarasingheAMY22} or studied compliance with respect to the privacy regulations of the website's country~\cite{DegelingULHSH19,OgutTMLYCU24}. Such studies require a preliminary website classification step to infer semantic information like the website category or country, and often rely on ad-hoc dataset construction techniques.

In general, labeling website datasets for targeted measurements is complex, costly, and error-prone. High-quality labels can be collected through human evaluators with sufficient domain expertise. Unfortunately, manual labeling has a significant cost, does not scale, and makes it difficult to expand or replicate existing studies. Automated labeling, in turn, is cheap and easy to scale, yet it typically relies on heuristics that can introduce bias or inaccuracies. For example, the country of a website can be inferred from its top-level domain, like .br or .de~\cite{OgutTMLYCU24}. This approach is useful but inferior to manual labeling, since it cannot label websites under generic top-level domains like .com or .net, which systematically biases study results, since such domains are common in practice.

Motivated by the explosive growth of generative AI, and by the many success stories of Large Language Models (LLMs) in particular, we explore the use of LLMs for creating labeled datasets of websites to enable targeted web measurements. The key intuition of our proposal is that LLMs can perform automated website classification by leveraging contextual information, natural language understanding, and the extensive knowledge they have gained from training on massive datasets constructed by scraping diverse web sources. An additional advantage of LLMs is their ability to adapt to different classification tasks through zero-shot or few-shot prompting, without requiring task-specific training or fine-tuning. In contrast, traditional machine-learning approaches typically require dedicated labeled training data, feature engineering, or task-specific model adaptation. The same general-purpose LLM can instead be repurposed across heterogeneous website classification tasks by specifying the target classes and the desired output through the prompt. This flexibility is extremely valuable for targeted web measurements, where the classes of interest may vary substantially across studies and developing a dedicated classifier for each measurement task would introduce additional engineering and data-collection costs. Together, these properties make LLMs particularly attractive compared with custom ad-hoc heuristics, e.g., based on selected keywords, as they combine automated classification with semantic reasoning and substantial flexibility across measurement tasks. 

\subsection*{Contributions}
In this paper, we make the following contributions:
\begin{enumerate}
    \item Based on a literature review, we identify key website classification tasks used in past web measurement studies and critically assess prior approaches to address them (Section~\ref{sec:survey}). We highlight significant limitations in existing methodologies and datasets, and we introduce a curated benchmark of high-quality labeled website datasets that enable a systematic evaluation of automated classification techniques (Section~\ref{sec:benchmark}).

    \item Using our curated datasets, we conduct a comprehensive evaluation of different open-source LLMs across multiple website classification tasks under two configurations (with and without website access). Our results show that LLMs can achieve strong performance across diverse scenarios, while revealing a delicate trade-off between classification accuracy and computational efficiency (Section~\ref{sec:classification}).

    \item We use LLMs to emulate the setup of a targeted web measurement by automatically detecting adult websites in a subset of the Tranco list, without any need for custom techniques for dataset construction. Combined with our benchmark evaluation, our case study shows that LLMs can enable high-quality targeted web measurements at scale directly over the Tranco list (Section~\ref{sec:measurement}).
\end{enumerate}

To support reproducible science, we release our datasets, prompts, code, and the results of our manual validation processes discussed in the paper~\cite{Artifacts}.
\section{Targeted Web Measurements and Website Classification}
\label{sec:survey}

Based on a literature review, we identify three relevant website classification tasks that recur in targeted web measurements, i.e., measurements focused on websites belonging to specific classes of interest. We then also review the third-party services that can be used to perform these classification tasks. Based on the results of our investigation, we identify relevant case studies for our paper and motivate the use of LLMs in supporting targeted web measurements.

\subsection{Targeted Web Measurements in the Literature}
\label{app:literature_review}
Website classification is a broad research area, appealing to different audiences and communities. Since the goal of our work is to explore the use of LLMs to support web measurements, we restrict the focus of our literature review to measurement studies and we identify those requiring (or making use of) some form of website classification. 
Our methodology consists of these steps:
\begin{enumerate}
    \item We extract from DBLP all the papers published from 2018 to 2025 at the major academic conferences in the following fields: computer security (IEEE S\&P, NDSS, ACM CCS, USENIX Security), privacy (PETS), Internet measurements (IMC), and the Web (WWW and WebSci).
    \item We filter papers so as to only keep those that are most likely to present a web measurement. To do this, we first identify as potential candidates all the papers including a case-insensitive match for the sub-strings ``web'' or ``measur'' in their title. When DBLP lists sessions or tracks for conferences, we also consider papers falling in the web security and web privacy categories as potential candidates. Finally, we read the abstracts of the candidate papers to identify those actually performing a web measurement.
    \item We inspect the matching papers to determine whether their web measurement involves any website classification step. 
\end{enumerate}

Our methodology identified 107 measurement papers published at the surveyed top venues, 47 of which perform some form of website classification in at least one experiment (44\%). At a high level, we observe that several papers rely on website classification to create new datasets, which are essential for drawing the primary conclusions of their study. For example, they are only interested in specific website categories~\cite{GotzeMISL22,SamarasingheAMY22,VallinaFGVA19} or they classify websites by country to check compliance against local privacy regulations~\cite{DegelingULHSH19,OgutTMLYCU24}. The other papers instead perform website classification as a complementary part of a broader analysis, e.g., they check whether specific website categories are correlated with the security or privacy aspects under study~\cite{DimovaGJ23,DrakonakisIP20,SquarcinaTVCM21}. Table~\ref{tab:classification} summarizes recurring website classification tasks identified in the surveyed web measurement literature, with a few representative papers for each task.

\begin{table*}[t]
\centering
\caption{Website classification tasks in web measurements.}
\label{tab:classification}
\begin{tabular}{@{}ll@{}}
\toprule
\textbf{Classification} & \textbf{Example Applications} \\ 
\midrule
\multirow{4}{*}{By category} & Analysis of governmental websites~\cite{GotzeMISL22,SamarasingheAMY22,SinganamallaJAK20} \\
& Privacy analysis of pornographic websites~\cite{VallinaFGVA19} \\
& Social studies and disinformation~\cite{SpangherRNFH20,ZengWGKR21} \\
& Breakdown results by website category~\cite{MendozaCG18,SquarcinaTVCM21} \\
\midrule
\multirow{2}{*}{By country} & Compliance with privacy regulations~\cite{DegelingULHSH19,OgutTMLYCU24} \\
& Socio-economic studies~\cite{BhuiyanVSZ25} \\
\midrule
\multirow{2}{*}{By functionality} & Identify websites with a private area~\cite{Al-Roomi023,Alroomi023,DrakonakisIP20} \\
& Identify websites with SSO access~\cite{ArdiC23,DimovaGJ23} \\
\bottomrule
\end{tabular}
\end{table*}

\subsection{Website Classification Tasks}
Based on our literature review, we select three relevant website classification tasks to consider in our evaluation of LLMs for targeted web measurements.

\subsubsection{Governmental Websites}
Prior research analyzed relevant privacy risks for citizens associated with e-government practices~\cite{GotzeMISL22,SamarasingheAMY22,SinganamallaJAK20}. We consider the automated detection of governmental websites as a first task to test the classification power of LLMs for multiple reasons. First, the importance of the topic: e-government is becoming more widespread nowadays, thus drawing attention from the research community in the last few years. Moreover, previous studies crucially relied on the creation of datasets of governmental websites, which can be effectively used as a starting point for a systematic evaluation of the classification performance of LLMs. Finally, the detection of governmental websites can be interpreted as a binary classification task (governmental vs. non-governmental), which is regarded as a baseline task in automated classification.

\revision{It is worth noting that constructing a dataset of governmental websites is challenging, particularly when aiming to avoid introducing unwanted biases, since governmental websites are not necessarily hosted under dedicated top-level domains such as .gov~\cite{GotzeMISL22}. Indeed, some countries do not even provide any dedicated governmental top-level domain, e.g., German governmental websites are hosted under the national .de top-level domain. Consequently, a simple match over the top-level domain does not generalize and the semantic understanding capabilities of LLMs can be valuable for identifying governmental websites on the Internet.} 

\subsubsection{Website Country}
Prior privacy studies analyzed website compliance with respect to local regulations~\cite{DegelingULHSH19,OgutTMLYCU24}, which requires linking websites to the country of their primary target audience (e.g., GDPR applies to sites offering services in the European Union). This naturally leads to a multiclass classification task of particular interest to the web privacy community.

Unfortunately, classifying websites based on the country of their primary target audience is far from a simple task and prior work leveraged ad-hoc heuristics, e.g., by inferring the country from the top-level domain of the analyzed websites~\cite{OgutTMLYCU24}. 
\revision{For instance, under this approach, \texttt{ebay.co.uk} is correctly labeled as a British website; however, no country can be assigned to websites hosted under generic top-level domains such as .com, thereby introducing a bias in the resulting dataset. LLMs can provide an effective tool for the automated identification of the country where a website primarily operates, even when the top-level domain of the website is uninformative.}

\subsubsection{Website Category}
Multiple studies rely on website categorization out of necessity (because they carry out targeted measurements, e.g.,~\cite{VallinaFGVA19,ZengWGKR21}) or just to provide complementary insights (they break down analysis results by website category, e.g.,~\cite{MendozaCG18,SquarcinaTVCM21}). Traditional approaches to website categorization broadly fall into two categories. On the one hand, we have annotated website datasets like DMOZ / Open Directory Project~\cite{dmoz-odp}, Yahoo Directory~\cite{yahoo-directory}, and the already mentioned Alexa ranking. Most of these datasets have been discontinued and are no longer maintained, hence they cannot be used to meaningfully categorize today's websites. On the other hand, we have online website classification services such as McAfee SiteAdvisor~\cite{mcafee-siteadvisor}, VirusTotal~\cite{virustotal}, and Cloudflare Radar~\cite{cloudflare-categorization}, which we review in Section~\ref{sec:commercial}. These commercial services normally require premium access or put restrictions in their terms of service that complicate their adoption at scale~\cite{VallinaPFPGBHTV20}. Additionally, they rely on fixed taxonomies that may provide insufficient granularity for specific web measurement studies. This state of affairs supports the case for LLMs as a convenient and widely available website categorization service, offering a great deal of flexibility with respect to the categories of interest.

\subsubsection{Exclusions}
According to our literature review (see Table~\ref{tab:classification}), our analysis covers classification tasks involving website categories and countries, but excludes the detection of specific website functionalities, such as private areas and SSO access. We do not address this task in this work, as it can be effectively automated using web crawlers designed to detect registration and login pages at scale~\cite{DrakonakisIP20,Al-Roomi023,Alroomi023,JannettMWMWM24}. While enhancing such crawlers with LLMs could improve their accuracy, this is beyond the scope of our study. Recent work on the use of LLMs for web crawling~\cite{StafeevRSKP25} may be inspiring for this line of research.

\subsection{Third-Party Website Classification Services}
\label{sec:commercial}
Several third-party services provide website classification functionality, which may suggest that automated website classification is a solved problem in practice. However, their practical suitability for large-scale targeted web measurements is less clear. We therefore review the current landscape of website classification services as of 2026. A comparison of popular services is reported in Table~\ref{tab:cat_services}.\footnote{Strictly speaking, several of these services expose website classification information as part of a broader product rather than operating solely as classification services. This distinction is immaterial to our analysis.} We consider all the services reported in the original survey by Vallina et al.~\cite{VallinaPFPGBHTV20}, plus additional services found through our literature review and targeted web searches. The first observation we make is that most services are not readily usable for large-scale academic web measurements, as they either do not provide free APIs, impose strict rate limits, or forbid automated access. In particular, we observe that just four website classification services offer free, publicly accessible APIs: Cloudflare Radar (country), Cloudflare Domain Threat Intelligence (category), the Google Topics API (category), and VirusTotal (category). 

The Google Topics API is a browser mechanism for interest-based advertising that relies on a lightweight on-device classifier to associate visited websites with a predefined taxonomy of topics. Although designed for advertising, the underlying classifier could in principle be repurposed to automatically categorize websites in web measurement studies. However, the Topics API has been deprecated and is scheduled for removal~\cite{googleTopicsAPI,googleTopicsAPIdepr1,googleTopicsAPIdepr2, googleTopicsAPIdepr3}, making it unsuitable as a sustainable solution for website categorization. Other free alternatives for website categorization include Cloudflare Domain Threat Intelligence and VirusTotal. Due to the restrictive rate limits imposed on its free tier, we estimate that classifying 10k websites using Cloudflare Domain Threat Intelligence would take roughly 3,000 days (around eight years), given its limit of 100 requests per month~\cite{cloudflare_api_limits}. The same task would take 20 days for VirusTotal, given its cap of 500 requests per day~\cite{virustotal_api_limits}. As a result, large-scale targeted web measurements are impractical using these free-tier services.

In the end, the only promising service for website classification these days appears to be Cloudflare Radar, which offers a free-to-use service for website country attribution without strict rate limits. We evaluate the performance of Cloudflare Radar on our curated dataset of websites later in the paper, revealing limited effectiveness for the task.

\begin{table*}[t]
\centering
\caption{Comparison of website classification services.}
\label{tab:cat_services}
\begin{tabular}{p{0.30\linewidth}lp{0.35\linewidth}}
\toprule
\textbf{Service} & \textbf{Classification Type} & \textbf{Main Limitations (as of 2026)} \\
\midrule
Bitdefender~\cite{bitdefender} & Category & No free tier for the API~\cite{bitdefender_price,bitdefender_api} \\
Cloudflare Radar~\cite{cloudflare-categorization} & Country & -- \\
Cloudflare Domain Threat Intelligence~\cite{cloudflare_api_categories} & Category & Free tier is limited to 100 requests per month~\cite{cloudflare_api_limits} \\
Dr.Web~\cite{drweb} & Category & No free tier for the API~\cite{drweb_price,drweb_api} \\
Forcepoint~\cite{forcepoint} & Category & No free tier for the API~\cite{forcepoint_price,forcepoint_api} \\
FortiGuard~\cite{FortiGuard} & Category & No free tier for the API~\cite{fortiguard_api_pricing_1,fortiguard_api_pricing_2,fortiguard_api_pricing_3} \\
Google Topics API~\cite{googleTopicsAPI} & Category & Deprecated and soon to be removed~\cite{googleTopicsAPIdepr1,googleTopicsAPIdepr2, googleTopicsAPIdepr3} \\
McAfee~\cite{mcafee-siteadvisor} & Category & Does not provide APIs, designed for occasional use, and scraping by automated programs is not allowed~\cite{mcafee-siteadvisor} \\
Similarweb~\cite{Similarweb} & Category \& Country & No free tier for the API~\cite{similarweb_api_pricing} \\
Symantec~\cite{symantec} & Category & No free tier for the API~\cite{symantec_price,symantec_api} \\
Trend Micro~\cite{trendmicro} & Category & No free tier for the API~\cite{trendmicro_price,trendmicro_api} \\
VirusTotal~\cite{virustotal} & Category & API's free tier is limited to 500 requests per day~\cite{virustotal_api_limits} \\
Webshrinker~\cite{Webshrinker} & Category & No free tier for the API~\cite{webshrinker_api_limits} \\
\bottomrule
\end{tabular}
\end{table*}

\section{Benchmark Datasets for Website Classification}
\label{sec:benchmark}

A systematic assessment of the performance of classifiers requires high-quality datasets that are correctly labeled and capture relevant aspects of the problem at hand. In this section, we create benchmark datasets for different website classification tasks, starting from an analysis of the state of the art. In particular, we start by rigorously assessing the quality of existing datasets and understanding the details of the underlying classification tasks, which is important both to construct benchmarks supporting a principled experimental evaluation and to design accurate prompts for LLMs. Starting from the state of the art and a careful analysis of its limitations, we explain how we improve on it to build our benchmark datasets. We then assess the quality of our datasets and discuss remaining limitations. Table~\ref{tab:datasets} reports key dataset statistics.
Since our analysis involves a fair share of manual validation, the online repository does not include just the final datasets, but also all the supporting evidence about their quality and the limitations of the state of the art reported in the paper.

\begin{table}[t]
\caption{Dataset statistics.}
\label{tab:datasets}
\centering
\begin{tabular}{lrr}
\toprule
\textbf{Dataset} & \textbf{\#Instances} & \textbf{\#Classes} \\
\midrule
Governmental & 3,166 & 2 \\
Countries & 7,780 & 10 \\
Categories & 3,785 & 13 \\
\bottomrule
\end{tabular}
\end{table}

\subsection{Governmental Websites}

\subsubsection{State of the Art}
\revision{To the best of our knowledge, prior work on privacy analysis of governmental websites produced three datasets~\cite{GotzeMISL22,SamarasingheAMY22,SinganamallaJAK20}, two of which are publicly available~\cite{GotzeMISL22,SinganamallaJAK20}. Despite differences in their construction, they all follow the same general approach and start from a seed list of known official government web pages from different countries. These seed lists are collected from trusted sources, filtered using strong signals (e.g., the .gov top-level domain), manually validated, or scrutinized using a combination of these methods. The seed lists are then used to discover additional links to ministries and agencies referenced within them using web crawling.}

To understand whether existing datasets already allow for a systematic evaluation of LLMs for governmental website detection, we conducted a preliminary quality analysis. We first reviewed the dataset by Gotze et al.~\cite{GotzeMISL22}. According to the original paper, their study focuses on websites that are ``associated with a domain that is registered and used by a national government''~\cite{GotzeMISL22}. Unfortunately, this definition does not seem to fully reflect the actual nature of the dataset, complicating validation. For example, the dataset includes \url{https://laeggs.com/}, which is the website of the Louisiana Egg Commission in the United States. It is not directly managed by the U.S. federal government, but it is operated by the Louisiana Department of Agriculture and Forestry, and serves an educational purpose by informing consumers about the nutritional value of eggs and egg products. Similar scenarios are common in the dataset: based on a random sampling of 200 websites and a manual inspection of them, we estimated that the definition of governmental website we started from covers just 138 websites (69\%).


We thus propose the following \emph{revised} definition of governmental website for label validation: ``a governmental website is an official online platform created and maintained by a government entity, or an organization significantly controlled or owned by a government. A primary goal of a governmental website must be to deliver government services, such as announcements, communication, exchange of information, and point of service to their citizens''. This definition extends the notion of governmental website to those websites that are managed by any entity with strong connections to a national government, while it enforces the additional restriction that the website must deliver services to the citizens. This choice aligns with the three aforementioned studies, which analyze privacy~\cite{GotzeMISL22,SamarasingheAMY22} and security~\cite{SinganamallaJAK20} risks of governmental websites serving as points of interaction with services for citizens.

To confirm the correctness of our revised definition, we re-examined the random sample of 200 websites from the dataset by Gotze et al. and manually confirmed that our definition correctly captures 183 of them (92\%). As for the 17 cases that were not yet captured by our revised definition, we observed that 16 are false positives of the original dataset, because these websites do not appear to be governmental in any substantive sense. This shows that the dataset by Gotze et al. largely reflects a meaningful definition of governmental website, with an estimated label precision of 92\%, however the number of errors therein is not negligible.

As for the dataset by Singanamalla et al.~\cite{SinganamallaJAK20}, we observe that it presents a very important difference with respect to the dataset by Gotze et al. that we first reviewed. In particular, around 98\% of the websites in the dataset by Singanamalla et al. are hosted under a dedicated top-level domain like .gov, which is a strong indicator that these websites are indeed governmental. This significantly differs from the distribution observed in the dataset by Gotze et al., where just 67\% of the websites are hosted under a dedicated top-level domain. We thus expect the dataset by Singanamalla et al. to have a high label precision, however it does not fully reflect the complexity of the classification task, because the dataset by Gotze et al. shows that governmental websites do not necessarily use dedicated top-level domains. Our assessment then reveals that existing datasets are sub-optimal, as they either contain a non-negligible number of false positives~\cite{GotzeMISL22} or are biased toward websites with dedicated top-level domains~\cite{SinganamallaJAK20}.

\subsubsection{Benchmark Construction}
To create a high-quality dataset with limited label noise, we first created a corpus including the 1,214 websites that belong to the intersection of the datasets by Gotze et al.~\cite{GotzeMISL22} and Singanamalla et al.~\cite{SinganamallaJAK20}. Since these websites are marked as governmental in two independent sources, we have high assurance about their governmental nature. Unfortunately, this construction alone would lead to a somewhat simplistic dataset, because the use of the intersection preserves the structural characteristics of the dataset by Singanamalla et al., leading to a dataset where around 99\% of the websites are hosted under a dedicated top-level domain. To compensate for that, we extended our initial corpus of 1,214 websites with 369 governmental websites that are not hosted under a dedicated top-level domain, leading to a final set of 1,583 governmental websites. Specifically, we extracted candidate websites from the dataset by Gotze et al. until we manually confirmed the governmental nature of 369 websites \revision{after visiting them}. This way, just 76\% of the websites in the extended corpus are hosted under a dedicated top-level domain, which makes the classification task more representative of the real-world complexity~\cite{GotzeMISL22}. 

Finally, to properly assess whether LLMs can actually distinguish governmental from non-governmental websites, we constructed a balanced dataset that assigns equal weight to both classes. In particular, we extended our set of 1,583 governmental websites with a random sample of 1,583 websites from the Tranco list that are not hosted under a known governmental top-level domain, e.g., .gov. To confirm the correctness of this random sampling, we accessed a subset of 200 websites from the sampled set and confirmed that 196 of them (98\%) were indeed non-governmental as expected. In the end, we obtained a reasonably sized dataset of 3,166 websites with negligible label noise thanks to the use of multiple independent sources and extensive manual validation.

\subsection{Website Country}
\label{sec:dataset_website_country}

\subsubsection{State of the Art}
We are not aware of any existing dataset associating websites with the country of their primary target audience, however there exist apparently promising techniques to consider to build such a dataset.
Ogut et al.~\cite{OgutTMLYCU24} proposed a simple heuristic to associate websites with the country of their primary target audience based on their top-level domain. This choice is reasonable and we can easily quantify its accuracy. By randomly sampling 200 websites from the Tranco list hosted under the country-specific top-level domains of the eight countries considered by Ogut et al.~\cite{OgutTMLYCU24}, we confirmed that 191 cases (96\%) were associated with the correct country based on their top-level domain alone. A major limitation of this approach is that websites hosted under generic top-level domains like .com and .net cannot be labeled using this simple methodology. This significantly complicates country attribution in large-scale measurements, e.g., over Tranco, where generic top-level domains are widespread, meaning that the technique by Ogut et al.~\cite{OgutTMLYCU24} cannot be used to build a dataset that actually reflects the complexity of the classification task.

We thus considered a different solution for country attribution based on the CrUX domain list~\cite{crux}. CrUX ranks domains based on the number of completed page loads observed among Chrome users and publishes country-specific top lists containing the domains most frequently accessed by users in each country. To assess whether membership in these lists provides a sufficiently discriminative signal for country attribution, we downloaded the top-10k lists for five diverse countries (China, Germany, Italy, Korea, and Turkey) and measured their overlap. If the lists captured substantially country-specific audiences, we would expect them to exhibit considerable differentiation. Instead, we observe that 4,017 domains appear in all five lists, corresponding to more than 40\% of each individual list. Given this significant overlap across countries, we then experimented with the following simple algorithm for country attribution: $(i)$ if a website belongs to a single country-specific list, we associate it with the label of that country; $(ii)$ if a website belongs to the intersection of all the lists, we mark it with the International label. Websites that meet neither criterion are left uncategorized, since we lack sufficient evidence to reliably label them.

To assess the quality of this labeling approach, we leverage the observation by Ogut et al.~\cite{OgutTMLYCU24} that the adoption of a country-specific top-level domain (TLD) is a strong predictor of a website's country, i.e., we can use this information to easily detect \emph{incorrect} labels. For example, given that a website hosted under the .it top-level domain is most likely Italian, we know that labeling such a website as German is most likely incorrect. In particular, we stipulate that:
\begin{itemize}
    \item A TLD is compatible with a country label if and only if it is the country's designated TLD or a generic TLD. For example, .de and .com are compatible with Germany, while .it and .fr are not (i.e., they are unlikely to host German websites);
    \item A TLD is compatible with the International label if and only if it is not country-specific. For example, .com and .net are compatible with the International label, while .tr and .jp are not (i.e., they are unlikely to host international websites).
\end{itemize}

As it turns out, the proposed preliminary labeling based on CrUX is potentially very imprecise, because it assigns a significant number of likely incorrect labels (we estimate at least 26\% incompatible cases for each of the labels). Full details are reported in Table~\ref{tab:preliminary-country}.

\begin{table}[t]
\caption{Preliminary country attribution attempt based on CrUX.}
\label{tab:preliminary-country}
\centering
\begin{tabular}{lrrr}
\toprule
\textbf{Label} & \textbf{\#Websites} & \textbf{\#Compatible TLD} & \textbf{Percentage} \\
\midrule
China & 3,910 & 2,890 & 74\% \\
Germany & 711 & 440 & 62\% \\
Italy & 301 & 135 & 45\% \\
Korea & 1,260 & 818 & 65\% \\
Turkey & 648 & 291 & 45\% \\
International & 4,017 & 2,828 & 70\% \\
\bottomrule
\end{tabular}
\end{table}

We then manually inspected 10 websites with a compatible top-level domain for each of the considered labels to understand whether filtering out websites hosted under incompatible top-level domains would be sufficient to perform country attribution with high precision. We managed to access 48 of the 60 sampled websites and manually verified that country attribution was correct in just 14 cases (29\%), i.e., removing websites hosted under incompatible top-level domains is insufficient for reliable country attribution. This suggests that country-specific lists from CrUX cannot be readily used on their own to solve the classification task at hand.

\subsubsection{Benchmark Construction}
The previous discussion showed that associating websites with the country of their primary target audience is a difficult task, despite the existence of country-specific top lists like those available in CrUX. Based on the analysis of the results of our manual investigation, we observe that many labeling errors can be readily fixed by complementing CrUX with additional signals. For example, checking the website language and the IP address of the host serving the content helps achieve reliable country attribution. In particular, we empirically observe the following facts:
\begin{enumerate}
    \item Local websites often serve content in the primary language of their country, irrespective of the language of the browser used to access them.
    \item International websites often adapt their content to the language of the requesting browser, based on the most prominent countries of their target audience.
    \item Local websites are often served by hosts operating in a nearby geolocation, e.g., a French website is unlikely to be hosted in Korea.
\end{enumerate}

These signals are not perfect. For example, international websites may just serve their content in English without any customization, and the language of local websites in English-speaking countries may be only a weak indicator of the website's country. While not fully general, these signals are valuable for our dataset construction, because they allow us to label a subset of websites in the CrUX lists with high accuracy. 

Given a set of candidate countries $C$, we then propose the following heuristic approach for country attribution:
\begin{enumerate}
    \item If a website belongs only to the top list of the country $c \in C$ and is hosted under the top-level domain of $c$, we immediately assign it label $c$. In this case, the top-level domain is already a strong indicator of the website's country~\cite{OgutTMLYCU24}, which is further supported by the presence of the website in a single country-specific top list.
    \item If a website belongs just to the top list of the country $c \in C$ and is hosted under a generic top-level domain, we access it with a browser set to English and we assign it label $c$ provided that $(i)$ the language of its homepage matches the primary language of $c$ and $(ii)$ the IP address of the host serving the content is in the same continent as $c$. In this case, we identify the country of a website based on its popularity within a single top list, supported by plausible language and geolocation information.
    \item If a website belongs to the intersection of all the lists and is hosted under a generic top-level domain, we access it multiple times with a browser set to the different languages of the countries in $C$ and we assign it the International label provided that the website localizes content to the requested language for at least three countries.
\end{enumerate}

Language attribution is performed using two Python libraries (fastText~\cite{fasttext} and gcld3~\cite{cld3}) and we consider language attribution successful if and only if at least one of the two libraries returns the official language of the country with probability at least 0.9. \revision{IP geolocation data is instead obtained through the IPinfo API~\cite{ipinfo_api}. We use this information to validate our classification and filter out false positives, since websites primarily targeting users in a given country seem unlikely to be hosted outside the same continent.}

We applied this construction process to a set of nine countries, including the five countries we experimented with (China, Germany, Italy, Korea, and Turkey) and four additional ones (India, Indonesia, Japan, and the Netherlands). This resulted in a dataset of 7,780 websites, including 630 websites serving an international audience and 7,150 local websites from nine countries. Out of the local websites, we observe that China is the country with the most websites served under a generic top-level domain (70.4\%), while India shows the opposite trend, with just 6.3\% of the websites in our dataset hosted outside the country's designated top-level domain.

To confirm the quality of our labeling process, we randomly sampled 20 websites for each of the considered countries and 20 international websites, leading to a set of 200 websites for label validation. Overall, the labeling process turned out to be very accurate upon manual validation, based on appropriate website interactions and the use of translation tools. Our manual validation discovered just 6 errors in total for automated labeling, i.e., 97\% of the labels in the sampled set are correct. We observe that 5 of the 6 errors occurred because a country label was assigned to a website actually serving an international audience and 3 of the 6 errors occurred on websites hosted under a country-dedicated top-level domain.

\subsection{Website Category}

\subsubsection{State of the Art}
We already mentioned that there exist a number of website classification services that may in principle be used to predict website categories, however they are not available at scale for different reasons (see Table~\ref{tab:cat_services}). A significant contribution in the field of website categorization is the curated snapshot of the Curlie dataset by Lugeon et al.~\cite{LugeonP022}. Curlie is a community-driven successor of DMOZ that inherits its taxonomy and URLs while providing updates, making it a strong starting point for dataset construction. Lugeon et al. curated a Curlie snapshot to improve data quality and label consistency, including uniformly translating top-level category names to English. This operation is fundamental for any rigorous evaluation based on the dataset.

The resulting dataset was also split into a training set and a test set of around 900k and 90k websites, respectively. The training set was used to learn a classifier for automated website categorization, called Homepage2Vec. However, Lugeon et al. found that Curlie is not exhaustively labeled, since contributors typically assign only a subset of relevant categories. Indeed, the classification task itself is very nuanced and multi-label in nature: for example, a sports magazine may be labeled as either Sports or News. This apparently conflicting information is not a mistake, just a different view of the primary purpose of the website. As a result, their curated dataset inevitably inherits the same limitation. To validate this, they crowdsourced the task of labeling a small dataset of 807 websites to human workers: while Curlie contained 836 labels, crowd workers produced 2,088 -- a 2.5x increase. This discrepancy highlights the need for a better dataset. Indeed, Homepage2Vec achieved significantly higher performance on the manually relabeled dataset than on the original test set, suggesting it often predicts all relevant classes even when ground truth labels are incomplete.

\subsubsection{Benchmark Construction}
In principle, thanks to its encouraging performance, we may use Homepage2Vec to mitigate the under-labeling problem of Curlie, e.g., by extending the labels of Curlie with those predicted by Homepage2Vec. However, we prefer avoiding the use of an automated classifier like Homepage2Vec in our dataset construction, because any (unavoidable) prediction error would lead to label noise. Our choice is to prioritize the information manually provided by the human annotators of Curlie to construct a more trustworthy dataset, mitigating the issues of the original curated snapshot by Lugeon et al.~\cite{LugeonP022}.

To do this, we first scraped fresh data from Curlie. This is useful to get access to additional metadata that we may use in our dataset construction, as well as to collect updated website categories, given that websites may be repurposed over time. In particular, contrary to what was done by Lugeon et al.~\cite{LugeonP022}, for each website we extracted the entire category tree as available in Curlie rather than just its top-level category (the root of the tree). For example, a website may be categorized as ``News'' / ``Media Industry'' / ``Services'' / ``Business Services for Media''. Here, the top-level category is News, but the category tree suggests that the website likely also belongs to the Business category: this is useful to mitigate under-labeling. 

A challenge here is that, while top-level categories belong to a small set of 13 known entries, sub-categories are manually created and maintained by human editors. Therefore, we apply tokenization and stemming, as provided by the NLTK library~\cite{nltk}, to each level of the category tree. For each website, we construct a set of candidate categories by matching the NLTK-processed output against the top-level categories, and by including the original top-level category itself. In our example, the website would be given two candidate categories: News (the top-level category) and Business (by tokenizing and stemming the string ``Business Services for Media'', to find the ``Business'' category). Each candidate category is then confirmed or dropped by applying a similar NLP-based approach over the website description as available in Curlie. Specifically, we match the NLTK-processed output of the website description against a manually curated dictionary that associates top-level categories with relevant keywords. In this case, if the website description reads ``Free press release distribution service that helps feed news.google.com, opt-in journalists, online media, freelance writers, and newspapers'', both candidate categories are confirmed thanks to the presence of the keywords ``press'' (associated with News) and ``service'' (associated with Business). This way, we use the entire category tree to extract additional candidate categories, but we assign categories only when we find additional evidence in the website description. The intuition is well-grounded in the official Curlie Editorial Guidelines, stating that descriptions should be ``concise, informative, and objective, telling end-users what they will find when they visit a website'' and must reflect the site's unique subject matter and content for correct categorization~\cite{curlieguidelines}. 



Using this approach, we created a dataset of 3,785 websites with their associated categories, and we estimate this dataset to be of higher quality than the original curated Curlie snapshot we started from~\cite{LugeonP022}. In particular, to confirm the quality of our labeling process, we performed a careful manual validation of a subset of the data. We randomly sampled 150 websites from our dataset, out of which 20 have more than one category based on our automated labeling process, and we manually reviewed them. The goal of our manual review was assigning all relevant Curlie categories to the sampled websites, following Curlie's editorial guidelines. We assessed the quality of the constructed dataset by comparing the automatically assigned labels of the 150 sampled websites against the manually assigned labels. As it turns out, 86 websites (57\%) have been assigned exactly the same labels in the two cases. As for the remaining 64 cases, we observe that 47 websites (31\%) have been automatically assigned a proper subset of the manually assigned labels, meaning that under-labeling is still present, yet most of the automatically assigned labels are correct, because they have also been given by human annotators. To get a better sense of these numbers, it is instructive to assess how the 150 considered websites have been originally annotated in the Curlie snapshot built by Lugeon et al.~\cite{LugeonP022}. A comparison with the manually annotated dataset shows that just 67 websites (45\%) have exactly the same labels. As for the remaining 83 cases, we notice that 56 websites (37\%) have been marked with a proper subset of the manually assigned labels. In short, our automated annotation procedure improved the quality of the dataset originally released by Lugeon et al., because it features more exact matches (+19) against our manually created dataset and reduces the under-labeled cases (-9). 

In the end, our analysis suggests that our automatically constructed dataset improves over a publicly available, state-of-the-art dataset. However, even our use of smarter and more effective heuristics cannot entirely compensate for the under-labeling phenomenon observed for the Curlie data source~\cite{LugeonP022}. 

\subsection{Limitations}

Despite our best efforts, the constructed datasets exhibit limitations, as their construction relies on heuristics motivated by the lack of a high-quality ground truth. Our manual validation confirmed that our datasets exceed the quality of state-of-the-art datasets used in prior work. Nevertheless, our datasets remain subject to specific issues and sources of bias. For instance, the Governmental dataset contains a large fraction of websites using dedicated top-level domains such as .gov, while only a smaller portion consists of governmental services hosted under generic domains such as .com. Although dedicated top-level domains are a strong indicator of governmental content, the true distribution of such domains across governmental websites on the Web is unknown. Similarly, the Countries dataset primarily includes websites from countries whose main language is predominantly spoken within national borders; thus, language becomes a strong predictive signal for country identification. This stems from the need to leverage reliable signals during our automated dataset construction. However, websites associated with multilingual countries or written in widely spoken languages (e.g., English, Spanish, or French) are likely more challenging to classify accurately, and may therefore provide a more informative testbed for evaluation. Finally, the Categories dataset is largely derived from Curlie, which is under-labeled and may not be fully representative of the broader Web.

Still, such limitations are inherent to the website classification problems at hand. In the absence of an independently available ground truth, constructing large-scale benchmark datasets necessarily requires relying on imperfect but high-confidence signals. Indeed, if website labels could already be obtained reliably through a general automated procedure, the classification problem addressed in this work would be largely solved. In the following, we therefore acknowledge the aforementioned limitations and use our benchmark datasets as a practical basis for evaluating LLMs on relevant website classification tasks. Their careful construction and manual validation provide a stronger basis for evaluation than existing datasets alone. We further complement these benchmark-based experiments with targeted analyses and more extensive experiments in the wild in Section~\ref{sec:measurement}.
\section{LLMs for Website Classification}
\label{sec:classification}

We here evaluate the performance of the tested LLMs on our benchmark datasets. Our evaluation considers two axes: $(i)$ \emph{classification performance}, i.e., measuring how accurately LLMs label websites across our three tasks (Section~\ref{sec:classification-performance}); and $(ii)$ \emph{computational cost}, i.e., measuring the time spent on the classification process and estimating its scalability (Section~\ref{sec:scalability}).

\subsection{Experimental Setup}
\label{sec:setup}
We here clarify the most important details of our experimental setup.

\subsubsection{Choice of the LLMs}
Nowadays, there are a plethora of LLMs available and their performance remains a subject of debate. In this work, we focus on self-hosted and open-source LLMs available in Ollama~\cite{ollama2025}. Although proprietary models such as ChatGPT or Gemini may achieve higher absolute performance, they rely on closed training pipelines and evolving inference-time optimizations. By focusing on open-source, self-hosted LLMs, we prioritize reproducibility and transparency in line with open science practices. Moreover, we are able to perform our experiments free of charge on our local infrastructure. This requirement is particularly important because web measurements are often large-scale, e.g., including over 100k websites, thus requiring multiple queries to LLMs. Finally, we expect self-hosted LLMs to be particularly appealing to security and privacy researchers, because they enable local computations over sensitive data. This mitigates ethical concerns when sensitive research data must not be shared with third parties.

\begin{table*}[t]
\centering
\small
\caption{Models considered for our experimental evaluation.}
\label{tab:model-selection}
\begin{tabular}{llcp{6cm}}
\toprule
\textbf{Model} & \textbf{Company} & \shortstack{\textbf{Included in}\\\textbf{Evaluation?}} & \textbf{Reason for Exclusion} \\
\midrule
deepseek-r1:70b         & DeepSeek    & No  & Last updated more than 1 year ago \\
gemma4:31b              & Google      & Yes & -- \\
gpt-oss:20b             & OpenAI      & Yes & -- \\
mistral-medium-3.5:128b & Mistral AI  & No  & Even the smallest variant cannot be fully loaded and executed on our GPU \\
muse-glimmer:30b        & Meta        & Yes & -- \\
nemotron3:33b           & NVIDIA      & Yes & -- \\
phi4-mini:3.8b          & Microsoft   & No  & Last updated more than 1 year ago \\
qwen3.8:27b             & Alibaba     & Yes & -- \\
\bottomrule
\end{tabular}
\end{table*}

We start from a list of recent general-purpose LLMs released by major companies and available in Ollama. We exclude models that: $(i)$ do not support the invocation of external tools, since we also evaluate their performance when given access to live websites through Playwright~\cite{playwright}; $(ii)$ have not been updated for more than one year as of August 27th, 2026; or $(iii)$ cannot be fully loaded into our NVIDIA A40-48Q GPU (48 GB VRAM) without CPU offloading. If there are multiple sizes available for a model that satisfies our criteria, we take the largest model that can be fully loaded and executed on our GPU without CPU offloading. Table~\ref{tab:model-selection} lists the considered candidates and the five models retained: gemma4:31b, gpt-oss:20b, muse-glimmer:30b, nemotron3:33b, qwen3.8:27b. These models are widely available and maintained by different companies, providing a meaningful basis for our investigation. 

\subsubsection{LLM Configuration and Prompt Design}
\label{sec:prompt_design}
We consider two LLM modes of operation: with and without live website access. In both modes, we use the default timeout of 600 seconds per URL: the LLM must return a prediction within this limit, regardless of any intermediate step performed. When a prediction attempt triggers the timeout, we consider it uncertain, because an answer may be returned with additional resources. Therefore, when we estimate classification performance, we present intervals for our performance measures, corresponding to all the uncertain predictions being assumed as always correct or always wrong. Access to live websites is mediated by the open-source LangChain framework~\cite{LangChain}, which enables integration with external tools such as the Playwright browser automation framework. We use LangChain to restrict the LLM to two actions: top-level browser navigation and extraction of the page's human-readable HTML text, executed through Playwright. These restrictions keep execution times reasonable while encouraging the model to rely on the semantic content of the visited page for classification. The actions are invoked within a multi-step inference pipeline where model reasoning may interleave with data retrieval, and the LLM autonomously decides whether to invoke a browser action or output a prediction directly. As a result, even with live access enabled, the model may classify a website without visiting it, e.g., if its internal knowledge is sufficient. This behavior is inherent to the pipeline and cannot be externally enforced, since the model may ignore website text even when provided in the prompt. Finally, we set the \emph{temperature} of all tested LLMs to 0 to reduce non-determinism and improve reproducibility, and we keep each model's thinking capabilities at the default Ollama configuration. To verify that this setting yields stable predictions, we conducted a preliminary experiment in which we compared the predictions across multiple runs and measured the standard deviation of our performance measures (see Appendix~\ref{app:temperature}).

Our prompt design follows established prompt engineering practices. We use \emph{persona assignment} to improve task adherence~\cite{ParkOCMLB23} and \emph{one-shot prompting}~\cite{BrownMRSKDNSSAA20} to increase response accuracy by providing a single, well-defined example of the desired output. 
For example, the prompt used to categorize governmental websites with live access enabled is the following:

\begin{tcolorbox}[colback=red!5!white,colframe=red!75!black,boxrule=0.5pt]
You are a classifier used to categorize websites into governmental and non-governmental websites.
A governmental website is an official online platform created and maintained by a government entity or an organization significantly controlled or owned by a government.
A primary goal of a governmental website must be to deliver government services, such as announcements, communication, exchange of information, and point of service to their citizens.
You are used by a research team conducting web measurements. You will be given one website:

\begin{itemize}
\item Identify whether the website is a governmental website or not based on our definition.
\item Do not modify the provided URL.
\item Visit the website in order to provide an accurate response.
\item Do not excessively rely on the .gov TLD: although this is likely a strong signal of governmental websites, some websites are operated by governments, but do not offer any services to citizens.
\end{itemize}

For example, https://www.pagopa.gov.it/ would be categorized as a governmental website, since it is run by the Italian government and allows citizens to perform online payments for governmental services.

Here is the website: \textbf{\textit{\textcolor{red}{<url>}}}.

Return a dictionary that tells us whether a URL is a governmental website, with two keys: URL and gov\_site.
Respond using JSON only.
\end{tcolorbox}

The prompt follows the same general structure as all other prompts in our study, making it easy to adapt to other website classification tasks. After persona assignment~\cite{ParkOCMLB23}, the prompt provides a precise definition of the classification task. This is followed by a set of instructions that help the model's reasoning process, along with an example for reference (one-shot prompting~\cite{BrownMRSKDNSSAA20}). To make results more reliable, we noticed that it is important to explicitly instruct the LLM not to modify the provided URL. Finally, the prompt includes the website (or list of websites) to be classified and specifies the required output format. In our prompts, the expected output is always a JSON dictionary containing two keys: the URL and the corresponding classification result. The full list of prompts used can be found in our online repository~\cite{Artifacts}.

\subsubsection{Performance Measures}
\label{app:perf_measures}
We note that our datasets differ in nature. The Governmental and the Countries datasets are single-label. For these datasets, we evaluate LLMs using two well-known, standard performance measures: \emph{accuracy} and \emph{macro F1 score}. Accuracy, defined as the ratio of correct predictions to the total number of predictions, provides an overall measure of classifier performance on balanced datasets. For unbalanced datasets like Countries, the F1 score is commonly used; it is defined as the harmonic mean of precision and recall, thereby combining both aspects into a single metric. The macro F1 score extends this concept to the multiclass classification setting by computing a per-class F1 score and averaging them, thus giving each class the same weight. This way, we can assess whether performance degrades substantially on specific classes, e.g., macro F1 penalizes models that are good at classifying German websites, but perform poorly on Italian websites. 

Choosing performance measures is more delicate for the Categories dataset, given its multi-label nature. In particular, we replace accuracy with \emph{Jaccard similarity}, which measures, for each instance, the overlap between the predicted and true label sets as the ratio of their intersection over their union. This provides a natural set-based notion of partial correctness and rewards meaningful overlaps even when predictions are incomplete. We do not report accuracy anymore, given the under-labeled nature of the Categories dataset. Since (exact-match) accuracy estimates the ratio of correct predictions, it would be overly penalizing in our multi-label setting with incomplete labels, as any missing label in the dataset may cause an otherwise reasonable prediction to be counted as incorrect.

\subsection{Classification Performance}
\label{sec:classification-performance}

\begin{table}[t]
\caption{LLM performance on the Governmental dataset. Best results for the two modes of operation are in bold, uncertain measures $(-)$ are associated with occasional timeouts during the classification process.}

\begin{subtable}[t]{\linewidth}
    \caption{Entire Governmental dataset.}
    \label{tab:performance-governmental}
    \centering
    \begin{tabular}{l|cc|cc}
    \toprule
    \multirow{2}{*}{\textbf{Model}}
        & \multicolumn{2}{c|}{\textit{URL only}} & \multicolumn{2}{c}{\textit{URL+Visit}}
        \\
        & Accuracy & Macro F1 & Accuracy & Macro F1
        \\
    \midrule
    gemma4:31b
        & 0.94--0.95 & 0.94--0.95 & \textbf{0.96--0.97} & \textbf{0.96--0.97}
        \\
    gpt-oss:20b
        & 0.92--0.93 & 0.92--0.93 & 0.95--0.96 & 0.95--0.96
        \\
    muse-glimmer:30b
        & \textbf{0.95} & \textbf{0.95} & \textbf{0.96--0.97} & \textbf{0.96--0.97}
        \\
    nemotron3:33b
        & 0.93 & 0.93 & 0.95--0.96 & 0.95--0.96
        \\
    qwen3.8:27b
        & 0.94 & 0.94 & 0.95--0.96 & 0.95--0.96
        \\
    \bottomrule
    \end{tabular}
\end{subtable}

\par\medskip

\begin{subtable}[t]{\linewidth}
    \caption{Subset of the Governmental dataset including just the URLs without a governmental TLD.}
    \label{tab:performance-governmental-non-gov-tld}
    \centering
    \begin{tabular}{l|cc|cc}
    \toprule
    \multirow{2}{*}{\textbf{Model}}
        & \multicolumn{2}{c|}{\textit{URL only}} & \multicolumn{2}{c}{\textit{URL+Visit}}
        \\
        & Accuracy & Macro F1 & Accuracy & Macro F1
        \\
    \midrule
    gemma4:31b
        & \textbf{0.92} & 0.87 & 0.94--0.95 & \textbf{0.91--0.92}
        \\
    gpt-oss:20b
        & 0.88--0.90 & 0.82--0.83 & 0.93 & 0.89--0.90
        \\
    muse-glimmer:30b
        & \textbf{0.92--0.93} & \textbf{0.88} & \textbf{0.95} & \textbf{0.91--0.92}
        \\
    nemotron3:33b
        & 0.89--0.90 & 0.81 & 0.93--0.94 & 0.89--0.90
        \\
    qwen3.8:27b
        & 0.91 & 0.84 & 0.93--0.94 & 0.89--0.90
        \\
    \bottomrule
    \end{tabular}
\end{subtable}
\end{table}

In the following we analyze the performance of the selected models over our three benchmark datasets.
 
\subsubsection{Governmental Websites}
Table~\ref{tab:performance-governmental} reports the accuracy and macro F1 scores computed for the different LLMs on the Governmental dataset. The results show that all models achieve near-optimal performance on the simplest of our classification tasks. Even without live website access, the worst-performing models reach an accuracy and F1 score around 0.93. When access to the websites is enabled, all models reach an accuracy and F1 score of at least 0.95, with a small improvement ranging between 0.01 and 0.03 depending on the model. 

Remarkably, this strong performance does not stem from the simple nature of the dataset alone, but is enabled by the use of powerful tools such as LLMs. Table~\ref{tab:performance-governmental-non-gov-tld} also reports our performance measures for the subset of the Governmental dataset comprising websites that are not hosted under a dedicated governmental TLD. For these websites, the URL provides a much weaker signal for classification purposes, and the LLMs have to rely more on their semantic understanding. Perhaps surprisingly, all models still show good to excellent performance on this more challenging subset of websites, even when they rely on the URL alone. The lowest estimated accuracy is 0.88 and the lowest estimated macro F1 score is 0.81. An important observation is that enabling live access to the websites substantially improves the classification performance of the models, with accuracy ranging from 0.93 to 0.95 and macro F1 scores ranging from 0.89 to 0.92. These figures provide clear evidence that all the evaluated state-of-the-art models can distinguish governmental websites from non-governmental websites very effectively, in particular when website access is enabled.

\begin{table}[t]
\caption{LLM performance on the Countries dataset. Best results for the two modes of operation are in bold, uncertain measures $(-)$ are associated with occasional timeouts during the classification process.}

\begin{subtable}[t]{\linewidth}
    \caption{Entire Countries dataset.}
    \label{tab:performance-countries}
    \centering
    \begin{tabular}{l|cc|cc}
    \toprule
    \multirow{2}{*}{\textbf{Model}}
        & \multicolumn{2}{c|}{\textit{URL only}} & \multicolumn{2}{c}{\textit{URL+Visit}}
        \\
        & Accuracy & Macro F1 & Accuracy & Macro F1
        \\
    \midrule
    gemma4:31b
        & \textbf{0.83--0.84} & \textbf{0.84--0.85} & \textbf{0.96} & \textbf{0.95}
        \\
    gpt-oss:20b
        & 0.81--0.82 & 0.80--0.82 & 0.94--0.95 & 0.92--0.94
        \\
    muse-glimmer:30b
        & \textbf{0.83} & 0.83--0.84 & \textbf{0.96} & 0.94--0.95
        \\
    nemotron3:33b
        & 0.79--0.80 & 0.79--0.83 & 0.89--0.90 & 0.86--0.89
        \\
    qwen3.8:27b
        & 0.81 & 0.82 & 0.95--0.96 & 0.92--0.95
        \\
    \bottomrule
    \end{tabular}
\end{subtable}

\par\medskip

\begin{subtable}[t]{\linewidth}
    \caption{Subset of the Countries dataset including just the URLs with a generic TLD.}
    \label{tab:performance-countries-int}
    \centering
    \begin{tabular}{l|cc|cc}
    \toprule
    \multirow{2}{*}{\textbf{Model}}
        & \multicolumn{2}{c|}{\textit{URL only}} & \multicolumn{2}{c}{\textit{URL+Visit}}
        \\
        & Accuracy & Macro F1 & Accuracy & Macro F1
        \\
    \midrule
    gemma4:31b
        & \textbf{0.52--0.53} & \textbf{0.50--0.51} & \textbf{0.90--0.91} & \textbf{0.85--0.87}
        \\
    gpt-oss:20b
        & 0.44--0.47 & 0.43--0.48 & 0.85--0.87 & 0.79--0.82
        \\
    muse-glimmer:30b
        & 0.50 & \textbf{0.50--0.52} & \textbf{0.90--0.91} & \textbf{0.85}
        \\
    nemotron3:33b
        & 0.39--0.43 & 0.42--0.49 & 0.69--0.72 & 0.67--0.72
        \\
    qwen3.8:27b
        & 0.46 & 0.47 & 0.89--0.90 & 0.83--0.85
        \\
    \bottomrule
    \end{tabular}
\end{subtable}

\end{table}

\subsubsection{Website Country}
Table~\ref{tab:performance-countries} reports the accuracy and macro F1 scores computed for the different LLMs on the Countries dataset. Performance measures are generally lower than those reported for the Governmental dataset, which is expected, because the Countries dataset has ten classes rather than two; this makes the classification task much harder. In light of this, one may argue that the LLM performance for website country attribution is indeed quite convincing: even without website access, accuracy ranges between 0.79 and 0.84 across models, while macro F1 scores range between 0.79 and 0.85. All models benefit from significant performance improvements when website access is activated. In this mode of operation, accuracy ranges between 0.89 and 0.96 across models, while macro F1 scores range between 0.86 and 0.95; improvements over the URL-only case are significant and may even reach 0.14, depending on the model and measure. 

Importantly, the LLMs with website access enabled also achieve good performance on the most challenging subset of the Countries dataset, which comprises websites hosted under generic TLDs such as .com and .net (Table~\ref{tab:performance-countries-int}). For these websites, the URL lacks a strong signal for classification purposes and no LLM achieves satisfactory performance without website access, with accuracy and macro F1 scores sitting below 0.55. This shows that website country attribution is already a complex task: although performance measures are good even when using URLs alone, they are influenced by the fact that national TLDs allow for a simple classification. However, once website access is enabled, most of the LLMs recover their predictive power. The best-performing models are gemma4:31b and muse-glimmer:30b, with accuracy around 0.90 and macro F1 score around 0.85. The worst model is nemotron3:33b, reaching at most 0.72 in both accuracy and macro F1 score; these numbers are significantly lower than those reported for the other models. 

For reference, we consider the Cloudflare Radar API as a possible alternative to LLMs for website country attribution, because this free service does not impose strict rate limits (see Table~\ref{tab:cat_services}). Cloudflare Radar determines website country attribution based on DNS queries observed by Cloudflare's infrastructure. Specifically, it returns a list of countries for each domain, along with the percentage of DNS requests originating from each country. After querying the API for all 7,780 websites in the Countries dataset, we found that Cloudflare Radar returned at least one country label for only 3,925 websites (50.4\%), indicating substantial coverage limitations. Unlike our ground truth labels, however, Cloudflare Radar does not provide an international label. To ensure a fair comparison, we therefore restricted the evaluation to websites whose ground truth label was not international and for which the most frequently reported country accounted for more than 50\% of DNS requests. This filtering resulted in a dataset of 2,344 websites, on which Cloudflare Radar correctly attributed the country in only 940 cases (40.1\%). These results suggest that Cloudflare Radar is not a suitable solution for large-scale website country attribution, due to both its limited coverage and low accuracy. Its reliance on DNS traffic observed by Cloudflare's infrastructure may help explain these limitations: websites generating little or no DNS traffic through Cloudflare may remain unlabeled or receive inaccurate country attributions. In contrast, LLMs label correctly around four out of five websites even when using their URL only and can also deal with complicated cases like generic TLDs when they are provided with website access.

\subsubsection{Website Category}
We last discuss the performance of LLMs for our most challenging task, i.e., the Categories dataset. Table~\ref{tab:performance-categories} reports Jaccard similarity and macro F1 scores for the different models. Performance measures appear modest in comparison with the previous tasks, in particular when LLMs have no website access. Even the best-performing model, muse-glimmer:30b, reaches at most 0.59 in Jaccard similarity and 0.55 in macro F1 score when using URLs only. Activating website access is very useful, boosting the performance measures of muse-glimmer:30b to 0.71 (+0.12) and 0.68 (+0.13) respectively. This is significantly better than competitors, because the second-best model gpt-oss:20b reaches at most 0.64 in Jaccard similarity and 0.63 in macro F1 score. These numbers are lower than those estimated for the previous tasks, also because the task itself is more nuanced and the quality of the ground truth is still not perfect, as we explained.

To put these numbers in context, we also evaluated the performance of Homepage2Vec, a state-of-the-art model for website classification trained over a snapshot of the Curlie dataset~\cite{LugeonP022}. Homepage2Vec shows a value of Jaccard similarity equal to 0.60 and a macro F1 score equal to 0.55 on the Categories dataset. This is quite worse than gpt-oss:20b and much worse than muse-glimmer:30b. This shows that general-purpose LLMs can label websites with higher quality than a classic machine learning model specifically trained for website categorization.

\begin{table}[t]
    \caption{LLM performance on the Categories dataset. Best results for the two modes of operation are in bold, uncertain measures $(-)$ are associated with occasional timeouts during the classification process.}
    \label{tab:performance-categories}
    \centering
    \begin{tabular}{l|cc|cc}
    \toprule
    \multirow{2}{*}{\textbf{Model}}
        & \multicolumn{2}{c|}{\textit{URL only}} & \multicolumn{2}{c}{\textit{URL+Visit}}
        \\
        & Jaccard & Macro F1 & Jaccard & Macro F1
        \\
    \midrule
    gemma4:31b
        & 0.44 & 0.48--0.49 & 0.56--0.57 & 0.60
        \\
    gpt-oss:20b
        & 0.51 & 0.49 & 0.64 & 0.63
        \\
    muse-glimmer:30b
        & \textbf{0.58--0.59} & \textbf{0.54--0.55} & \textbf{0.71} & \textbf{0.68}
        \\
    nemotron3:33b
        & 0.52--0.55 & 0.50--0.52 & 0.62--0.63 & 0.59
        \\
    qwen3.8:27b
        & 0.47 & 0.48 & 0.61--0.63 & 0.61--0.63
        \\
    \bottomrule
    \end{tabular}
\end{table}

\subsection{Computational Cost}
\label{sec:scalability}
Table~\ref{tab:model-times} reports the total, mean, and median classification times for all models and datasets. When live access to the website is enabled, the table also reports the time spent by the LLM and the time spent by browser access in separate columns. The table shows that mean classification times range from 3.05 to 39.07 seconds per website, depending on the model, dataset, and mode of operation. In all cases, the median is lower than the mean, suggesting that classification times are generally shorter than what is captured by the mean. The median can be much smaller than the mean when classification is based on URLs alone, reflecting that many websites can be classified more efficiently than what the mean suggests. When live access to websites is enabled, the difference between mean and median is relatively small in general, suggesting that the mean is more representative of the typical classification time.

Enabling live website access increases classification time in 10 of the 15 configurations, with increases ranging from 7.7\% (gemma4:31b on Countries) to 376.1\% (nemotron3:33b on Governmental), corresponding to factors of 1.08 to 4.76 relative to URL-only classification. Nevertheless, mean classification times with live access remain between 7.53 and 39.04 seconds per website, suggesting that website visits remain computationally feasible within the evaluated setting. Interestingly, gpt-oss:20b is the model that is consistently the most efficient when website access is enabled. Indeed, live website access consistently reduces total classification time for this model across all three datasets, yielding reductions of 30.9--32.5\%. To explain this, we observe that the time spent by the LLM decreases by 56.9--67.4\% when website access is activated, more than compensating for the additional browser access time. This pattern may be associated with the website content reducing the reasoning effort required for classification.

As for scalability, it is instructive to compare a few representative configurations:
\begin{itemize}
    \item nemotron3:33b takes around 3 seconds on average to detect governmental websites in URL-only mode. This is the fastest scenario, where nemotron3:33b achieves excellent performance (accuracy and F1 score of 0.93). This means that 10k websites can be classified as governmental or not in around 8 hours using this model on our hardware.

    \item gpt-oss:20b takes around 11 seconds on average to perform country attribution after website access. For this task, gpt-oss:20b is the fastest model and achieves excellent performance (accuracy around 0.94, F1 score around 0.92, with good results also on websites with generic TLDs). This means that the country of 10k websites can be predicted in around 30 hours using this model on our hardware.

    \item muse-glimmer:30b takes around 39 seconds on average to categorize websites after website access, representing the highest computational cost in our setting. For this task, this is the only configuration yielding good performance (Jaccard similarity 0.71, macro F1 score 0.68). This implies that the categories of 10k websites can be predicted in around 108 hours (4.5 days) using this model on our hardware. 
\end{itemize}

This suggests that using LLMs for website classification at scale is feasible, because labeling 10k websites takes from a few hours to a few days. Importantly, the labeling process is parallelizable: if one had $n$ GPUs, they could achieve a speedup of $n$ by breaking websites into disjoint batches and assigning them to different GPUs.

\begin{table}[t]
\caption{Total, mean, and median classification times of the LLMs on the different datasets. Best results for total, mean, and median classification times are in bold. When live access to the website is enabled, the table also reports the time spent by the LLM and the time spent by browser access in separate columns.}
\label{tab:model-times}

\begin{subtable}[t]{\linewidth}
    \caption{Results for the Governmental dataset.}
    \label{tab:model-tool-times-governmental}
    \centering
    \begin{tabular}{l|ccc|ccccc}
    \toprule
    \textbf{Model}
        & \multicolumn{3}{c|}{\textit{URL only}}
        & \multicolumn{5}{c}{\textit{URL+Visit}} \\
        & Total & Mean & Median
        & Total & Mean & Median & LLM & Browser \\
        & (h) & (s) & (s)
        & (h) & (s) & (s) & (h) & (h) \\
    \midrule
    gemma4:31b
        & 18.97 & 21.58 & 18.82
        & 28.60 & 32.52 & 29.12 & 24.66 & 3.94 \\
    gpt-oss:20b
        & 14.14 & 16.08 & 2.50
        & \textbf{9.77} & \textbf{11.11} & \textbf{7.11} & 5.16 & 4.61 \\
    muse-glimmer:30b
        & 22.28 & 25.33 & 19.56
        & 30.13 & 34.26 & 28.29 & 25.89 & 4.24 \\
    nemotron3:33b
        & \textbf{2.68} & \textbf{3.05} & \textbf{2.07}
        & 12.76 & 14.51 & 13.07 & 8.96 & 3.80 \\
    qwen3.8:27b
        & 7.85 & 8.92 & 8.15
        & 12.83 & 14.58 & 11.73 & 8.80 & 4.03 \\
    \bottomrule
    \end{tabular}
\end{subtable}

\par\medskip

\begin{subtable}[t]{\linewidth}
    \caption{Results for the Countries dataset.}
    \label{tab:model-tool-times-countries}
    \centering
    \begin{tabular}{l|ccc|ccccc}
    \toprule
    \textbf{Model}
        & \multicolumn{3}{c|}{\textit{URL only}}
        & \multicolumn{5}{c}{\textit{URL+Visit}} \\
        & Total & Mean & Median
        & Total & Mean & Median & LLM & Browser \\
        & (h) & (s) & (s)
        & (h) & (s) & (s) & (h) & (h) \\
    \midrule
    gemma4:31b
        & 45.40 & 21.01 & 16.29
        & 48.90 & 22.63 & 21.40 & 41.52 & 7.38 \\
    gpt-oss:20b
        & 34.43 & 15.93 & \textbf{2.16}
        & \textbf{23.26} & \textbf{10.76} & \textbf{7.45} & 11.21 & 12.05 \\
    muse-glimmer:30b
        & 60.87 & 28.17 & 22.76
        & 74.53 & 34.49 & 31.42 & 59.80 & 14.73 \\
    nemotron3:33b
        & 25.03 & 11.58 & 2.39
        & 38.93 & 18.01 & 12.25 & 21.49 & 17.43 \\
    qwen3.8:27b
        & \textbf{13.53} & \textbf{6.26} & 5.01
        & 39.50 & 18.28 & 16.42 & 25.22 & 14.28 \\
    \bottomrule
    \end{tabular}
\end{subtable}

\par\medskip

\begin{subtable}[t]{\linewidth}
    \caption{Results for the Categories dataset.}
    \label{tab:model-tool-times-categories}
    \centering
    \begin{tabular}{l|ccc|ccccc}
    \toprule
    \textbf{Model}
        & \multicolumn{3}{c|}{\textit{URL only}}
        & \multicolumn{5}{c}{\textit{URL+Visit}} \\
        & Total & Mean & Median
        & Total & Mean & Median & LLM & Browser \\
        & (h) & (s) & (s)
        & (h) & (s) & (s) & (h) & (h) \\
    \midrule
    gemma4:31b
        & 33.88 & 32.22 & 27.53
        & 37.60 & 35.76 & 31.59 & 35.34 & 2.26 \\
    gpt-oss:20b
        & 11.73 & 11.15 & \textbf{2.76}
        & \textbf{7.92} & \textbf{7.53} & \textbf{6.57} & 5.06 & 2.86 \\
    muse-glimmer:30b
        & 41.07 & 39.07 & 32.32
        & 41.05 & 39.04 & 37.20 & 37.87 & 3.18 \\
    nemotron3:33b
        & 21.07 & 20.04 & 3.83
        & 15.92 & 15.14 & 11.73 & 11.95 & 3.97 \\
    qwen3.8:27b
        & \textbf{9.47} & \textbf{9.00} & 8.73
        & 15.84 & 15.06 & 13.90 & 12.76 & 3.08 \\
    \bottomrule
    \end{tabular}
\end{subtable}
\end{table}

\subsection{Takeaways of the Evaluation}
We here clarify the key insights supported by our experimental evaluation, explaining how they can be useful to the web measurement community. LLMs are generally effective at website classification tasks, however the choice of the model and its mode of operation play a fundamental role in performance. Simple tasks like detecting governmental websites can be solved by all the tested models using just the URL of the website. More complex tasks, instead, fundamentally require live website access to be effectively solved with the evaluated models. We showed that both country attribution for websites hosted under a generic TLD and website categorization are already complex tasks that cannot be solved without website access. When the classification task becomes harder, the choice of the specific LLM becomes more important. For example, muse-glimmer:30b is the only tested LLM that can achieve satisfactory performance on the Categories dataset.

In general, requiring website access may increase the total classification time, however the average increase does not seem to fundamentally hinder scalability. For some models like gpt-oss:20b, enabling website access may even \emph{reduce} the total classification time, because the reduction in the time spent in LLM processing is significantly larger than the overhead introduced by website access. When evaluating different LLMs for adoption, this aspect should be carefully taken into account. In our experiments, muse-glimmer:30b is the slowest among all the tested models, with mean classification times ranging between 25 and 40 seconds depending on the task and mode of operation. In contrast, gpt-oss:20b is consistently the fastest model when website access is enabled, with mean classification times ranging between 7 and 11 seconds depending on the task, i.e., a 3x to 5x speedup with respect to muse-glimmer:30b. Although gpt-oss:20b is three to five times faster than muse-glimmer:30b, it also performs worse on all the considered classification tasks. This suggests a trade-off between classification performance and computational efficiency.
\section{Case Study: Enabling Web Measurements on Adult Websites}
\label{sec:measurement}
Our previous evaluation showed that LLMs enable effective website classification with an acceptable computational cost. We here move beyond the datasets used in our controlled evaluation to simulate the setup for a targeted web measurement on popular adult websites. 

\subsection{Experimental Setup}
Web measurements are typically conducted using the Tranco list, which has become the de facto standard in the field and has been adopted by more than 600 academic publications. However, Tranco is unannotated. As a result, measurements targeting specific classes of websites require ad-hoc filtering or heuristic-based dataset construction methods to isolate the domains of interest. To assess whether LLMs can support the construction of accurately labeled datasets for targeted web measurements, we consider adult websites as a case study. Previous work~\cite{VallinaFGVA19} on measuring privacy risks in pornographic websites constructed their dataset through three heuristics: $(i)$ filter the Alexa Top list to retain only URLs containing keywords strongly indicative of pornographic content (e.g., ``porn'' and ``sex''), $(ii)$ scrape websites specialized in aggregating, recommending, and classifying pornographic content, and $(iii)$ include domains categorized as ``Adult'' by Alexa's website classification service.

Instead, we build our dataset from Tranco using LLMs for website classification. In particular, we label the top 10k websites of Tranco as either pornographic or not using both gpt-oss:20b and muse-glimmer:30b, with live website access enabled. This significantly simplifies the dataset construction process. We consider gpt-oss:20b because it is the fastest model in this mode of operation and muse-glimmer:30b because it is the model generally providing the best classification performance. This way, we are able to evaluate two representative configurations: one with efficiency in mind and the other with optimal classification performance as the key target. We enable website access because our previous evaluation shows that it generally improves classification performance, speeds up classification with gpt-oss:20b, and introduces just a limited overhead for muse-glimmer:30b.

\subsection{Classification Performance}
Table~\ref{tab:confusion} compares the predictions of gpt-oss:20b and muse-glimmer:30b over Tranco Top 10k. The table shows that the performed predictions are largely in agreement: the two models return a different output in just 151 cases, i.e., their agreement rate is around 98.5\%. We manually confirmed that agreement is strongly correlated with correct predictions by sampling 50 websites marked as adult by both models and 150 websites marked as non-adult by both models. In particular, we successfully accessed the 200 websites using a standard web browser and we identified a single prediction error, i.e., a website incorrectly marked as adult by both models. This suggests that, when the two models are in agreement, their prediction is most likely correct.

We then tried to access all the 151 websites where gpt-oss:20b and muse-glimmer:30b returned different predictions, identifying 117 websites that provided a successful response. There we observed that muse-glimmer:30b performed the correct class prediction in 94 out of 117 cases (80.3\%), which confirms its higher predictive power with respect to gpt-oss:20b. In particular, we observed the following breakdown:
\begin{itemize}
    \item 31 websites have been correctly classified as adult by muse-glimmer:30b, while they have been incorrectly marked as non-adult by gpt-oss:20b.
    \item 63 websites have been correctly classified as non-adult by muse-glimmer:30b, while they have been incorrectly marked as adult by gpt-oss:20b.
    \item 10 websites have been correctly classified as adult by gpt-oss:20b, while they have been incorrectly marked as non-adult by muse-glimmer:30b.
    \item 13 websites have been correctly classified as non-adult by gpt-oss:20b, while they have been incorrectly marked as adult by muse-glimmer:30b.
\end{itemize}

This analysis shows that the most common type of error in the analyzed subset of websites is over-reporting from gpt-oss:20b.

\begin{table}[t]
    \caption{Comparison of the predictions of gpt-oss:20b and muse-glimmer:30b over Tranco Top 10k.}
    \label{tab:confusion}
    \centering
    \begin{tabular}{ll|cc}
    \toprule
        & & \multicolumn{2}{c}{\textbf{muse-glimmer:30b}} \\
    \cmidrule(lr){3-4}
        & & Adult & Non-Adult \\
    \midrule
    \multirow{2}{*}{\textbf{gpt-oss:20b}}
        & Adult    & 482 & 101 \\
        & Non-Adult & 50 & 9,367 \\
    \bottomrule
    \end{tabular}
\end{table}

To account for the oversampling of disagreements in our manual audit, we estimate classification performance using the four prediction strata in Table~\ref{tab:confusion}. The population sizes are $(482, 101, 50, 9367)$, their manually validated sample sizes are $(50, 73, 44, 150)$, and the numbers of manually confirmed adult websites are $(49, 10, 31, 0)$, respectively. We construct a weighted confusion matrix for each model by assigning every verified website in stratum $s$ a weight $N_s/n_s$, where $N_s$ and $n_s$ denote the population and validated sample sizes of that stratum. The resulting estimates, reported in Table~\ref{tab:weighted-performance}, indicate high overall accuracy for both models, with a substantially higher macro F1 score for muse-glimmer:30b. Note that macro F1 is much more important than accuracy in this setting, due to class imbalance. These estimates assume that the manually verified websites are representative of their respective prediction strata, including websites that could not be accessed. The stratum of 9,367 websites classified as non-adult by both models remains a source of uncertainty. Although no errors were observed among the 150 manually verified websites, shared false negatives among the unverified websites cannot be ruled out in general, given the size of the population.

\begin{table}[t]
    \caption{Estimated classification performance on Tranco Top 10k after stratified weighting of the manual audit.}
    \label{tab:weighted-performance}
    \centering
    \begin{tabular}{l|cc}
    \toprule
    \textbf{Model}
        & Accuracy & Macro F1 \\
    \midrule
    gpt-oss:20b
        & 0.987 & 0.937 \\
    muse-glimmer:30b
        & 0.996 & 0.981 \\
    \bottomrule
    \end{tabular}
\end{table}

\subsection{Computational Cost}
Table~\ref{tab:times-adult} reports the total, mean, and median classification times of gpt-oss:20b and muse-glimmer:30b over Tranco Top 10k. Labeling 10k websites with gpt-oss:20b takes around 10 hours on our hardware, while muse-glimmer:30b is significantly slower at this task and takes around 56 hours. If the performance figures generalize to different websites, gpt-oss:20b could label 100k websites in around four days, while muse-glimmer:30b would take around 23 days. This suggests that gpt-oss:20b can enable large-scale measurements, while the scalability of muse-glimmer:30b is more at risk. However, it is possible to combine the two models to achieve a better trade-off between classification performance and computational cost.

Since we showed that gpt-oss:20b is most often in agreement with muse-glimmer:30b, yet more susceptible to false positives, we estimate the computational cost of a two-step pipeline where gpt-oss:20b is first used to build a set of candidate adult websites and muse-glimmer:30b is just executed on the identified candidates to filter out false positives. This pipeline may miss some adult websites, because gpt-oss:20b also suffers from more false negatives than muse-glimmer:30b, however it is much more efficient than just running muse-glimmer:30b on the entire Tranco list and may be an acceptable practice when the web measurement is just concerned with covering a subset of adult websites, rather than their entire population. The classification times of the combined pipeline are reported in the last row of Table~\ref{tab:times-adult}. Labeling 10k websites with the proposed two-step pipeline takes around 14 hours, which is a +37\% overhead with respect to gpt-oss:20b, but just 25\% of the time required by muse-glimmer:30b. With this setup, we estimate that labeling 100k websites would take less than six days, which is appropriate for large-scale measurements. In terms of classification performance, based on our manually verified sample of websites, the two-step pipeline would filter out 63 websites that have been correctly classified as non-adult by muse-glimmer:30b alone, at the price of losing 10 websites that have been correctly classified as adult by gpt-oss:20b alone.

\begin{table}[t]
    \caption{Total, mean, and median classification times of gpt-oss:20b and muse-glimmer:30b over Tranco Top 10k. Best results for total, mean, and median classification times are in bold. The table also reports the time spent by the LLM and the time spent by browser access in separate columns. The \textit{combined pipeline} row reports results for the proposed two-step pipeline leveraging both models.}
    \label{tab:times-adult}
    \centering
    \begin{tabular}{l|ccccc}
    \toprule
    \textbf{Model}
        & \multicolumn{5}{c}{\textit{URL+Visit}} \\
        & Total & Mean & Median & LLM & Browser \\
        & (h) & (s) & (s) & (h) & (h) \\
    \midrule
    gpt-oss:20b
        & \textbf{10.21} & \textbf{3.67} & \textbf{1.36} & 6.32 & 3.89 \\
    muse-glimmer:30b
        & 56.22 & 20.24 & 19.60 & 46.87 & 9.36 \\
    \textit{combined pipeline}
        & 13.97 & 5.03 & \textbf{1.36} & 9.77 & 4.20 \\
    \bottomrule
    \end{tabular}
\end{table}

\section{Related Work}

Web measurements are popular in the security and privacy communities to understand the current state of the Web. Security measurements have assessed the adoption and configuration of important HTTP headers, like Content Security Policy~\cite{CalzavaraRB18,WeichselbaumSLJ16}, HTTP Strict Transport Security~\cite{KranchB15}, X-Frame-Options~\cite{CalzavaraRR0S20}, and Cross-Origin Resource Sharing~\cite{ChenJDWCP018}. Other work instead measured the prevalence and impact of significant web vulnerabilities, such as cross-site scripting~\cite{LekiesSJ13}, cross-site request forgery~\cite{SudhodananCCDAM17}, and web cache poisoning~\cite{Liang0GSJHYD24}. In the privacy field, web measurements largely focused on cookies~\cite{AcarEEJND14,EnglehardtN16}, browser fingerprinting~\cite{LaperdrixBBA20} and compliance with privacy regulations~\cite{DegelingULHSH19,OgutTMLYCU24}. As noted in Table~\ref{tab:classification}, several measurement studies involved some form of website classification either as a core component of the study or to provide a complementary perspective on specific website categories. Despite the importance of website classification, most previous work relied on ad-hoc heuristics that are imperfect and may unnecessarily bias research inferences towards websites amenable to classification. \revision{In this work, we argue that LLMs can serve as general and effective tools for website classification in web measurement studies. To support this claim, we systematically evaluate their performance and scalability and demonstrate their practical applicability in a targeted web measurement study over the Tranco list.} We are not aware of any systematic study on how LLMs can serve web measurements at this time.

Website classification is, in general, an important task~\cite{QiD09}, appealing to different communities. While some research focuses on general classification tasks~\cite{LugeonP022,YeBOATPA24}, most works in the web security field focus on the detection of malicious activity, such as fraudulent e-commerce pages~\cite{BitaabCOLWAWBSD23}, phishing websites~\cite{LinLDNCLSZD21}, and other types of malicious web pages. The main goal of this line of work is to improve the performance of existing detection approaches, rather than measuring the prevalence of malicious websites in the wild. We expect that LLMs can be successfully applied to this field as well, however our paper focuses on investigating the use of LLMs as a useful support for web measurement studies.

Finally, LLMs have been recently applied to web crawling~\cite{StafeevRSKP25}. In this field, LLMs are used to process web pages after extracting semantic information from them, to improve crawling coverage, and to trigger complex interactions. This ability of LLMs can certainly be useful in web measurements, given the importance of the crawler on research inferences~\cite{AhmadDZVN20}. \revision{However, this line of research is orthogonal to our work, which focuses on evaluating the ability of LLMs to perform effective website classification and on how they can support large-scale targeted web measurements.}
\section{Conclusion}
In this paper, we investigated the use of LLMs to support targeted web measurements through automated website classification. Building on a review of prior measurement studies, we identified common classification tasks and constructed curated datasets to systematically evaluate multiple LLMs with and without live website access. Our results show that LLMs can achieve strong classification performance at a computational cost compatible with large-scale measurements, while model choice and configuration substantially affect both quality and efficiency. Our case study on adult website identification in the Tranco list~\cite{PochatGTKJ19} extends this evaluation to the construction of a labeled dataset for a concrete measurement scenario. The benchmark experiments and the case study provide complementary evidence that LLMs can support accurate and scalable website classification, reducing reliance on manual labeling and task-specific heuristics when constructing datasets for targeted web measurements.

In future work, we plan to investigate how LLMs can help identify \emph{relevant} websites and web pages for specific security and privacy measurements. In particular, we aim to explore whether their semantic understanding can guide data collection toward relevant content, for example by filtering out mock websites and error pages. We also intend to evaluate how classification performance and computational cost are affected by the use of proprietary LLMs, such as Gemini, as well as fine-tuning~\cite{WangCJPCYY25}, in-context learning~\cite{Dong0DZMLXX0C0S24}, and different prompt designs or reasoning settings, particularly for more challenging classification tasks.

\bibliographystyle{ACM-Reference-Format}
\bibliography{biblio}

\appendix 
\section{Run-to-Run Variability in Classification Performance}
\label{app:temperature}
Table~\ref{tab:std-dev} reports the standard deviation of the performance measures of the LLMs over five runs on the different datasets. For performance reasons, each run operates on a sample of the original dataset, including 10\% of the instances and reflecting the same label distribution. The numbers show that setting the temperature to 0 ensures that the performance of the LLMs is stable across experiments, because the standard deviation never exceeds 0.008 and is often very close to 0.

\begin{table}[h]
\caption{Standard deviation of the LLM performance over five runs on the different datasets.}
\label{tab:std-dev}

\begin{subtable}[t]{\linewidth}
    \caption{Governmental dataset.}
    \label{tab:performance-std-governmental}
    \centering
    \begin{tabular}{l|cc|cc}
    \toprule
    \multirow{2}{*}{\textbf{Model}}
        & \multicolumn{2}{c|}{\textit{URL only}} & \multicolumn{2}{c}{\textit{URL+Visit}}
        \\
        & Accuracy & Macro F1 & Accuracy & Macro F1
        \\
    \midrule
    gemma4:31b
        & 0.000 & 0.000 & 0.000 & 0.000
        \\
    gpt-oss:20b
        & 0.004 & 0.004 & 0.005 & 0.005
        \\
    muse-glimmer:30b
        & 0.000 & 0.000 & 0.003 & 0.003
        \\
    nemotron3:33b
        & 0.000 & 0.000 & 0.002 & 0.002
        \\
    qwen3.8:27b
        & 0.000 & 0.000 & 0.001 & 0.001
        \\
    \bottomrule
    \end{tabular}
\end{subtable}

\par\medskip

\begin{subtable}[t]{\linewidth}
    \caption{Countries dataset.}
    \label{tab:performance-std-countries}
    \centering
    \begin{tabular}{l|cc|cc}
    \toprule
    \multirow{2}{*}{\textbf{Model}}
        & \multicolumn{2}{c|}{\textit{URL only}} & \multicolumn{2}{c}{\textit{URL+Visit}}
        \\
        & Accuracy & Macro F1 & Accuracy & Macro F1
        \\
    \midrule
    gemma4:31b
        & {0.000} & {0.000} & {0.001} & {0.002}
        \\
    gpt-oss:20b
        & 0.004 & 0.003 & 0.007 & 0.008
        \\
    muse-glimmer:30b
        & 0.003 & 0.003 & 0.004 & 0.003
        \\
    nemotron3:33b
        & {0.000} & {0.000} & {0.001} & 0.004
        \\
    qwen3.8:27b
        & {0.000} & {0.000} & 0.003 & 0.003
        \\
    \bottomrule
    \end{tabular}
\end{subtable}

\par\medskip

\begin{subtable}[t]{\linewidth}
    \caption{Categories dataset.}
    \label{tab:performance-std-categories}
    \centering
    \begin{tabular}{l|cc|cc}
    \toprule
    \multirow{2}{*}{\textbf{Model}}
        & \multicolumn{2}{c|}{\textit{URL only}} & \multicolumn{2}{c}{\textit{URL+Visit}}
        \\
        & Jaccard & Macro F1 & Jaccard & Macro F1
        \\
    \midrule
    gemma4:31b
        & 0.002 & 0.003 & 0.006 & {0.001}
        \\
    gpt-oss:20b
        & {0.000} & {0.000} & 0.003 & 0.005
        \\
    muse-glimmer:30b
        & {0.000} & {0.000} & 0.006 & 0.006
        \\
    nemotron3:33b
        & {0.000} & {0.000} & {0.001} & 0.003
        \\
    qwen3.8:27b
        & {0.000} & {0.000} & {0.001} & 0.002
        \\
    \bottomrule
    \end{tabular}
\end{subtable}

\end{table}

\end{document}